\documentclass[twocolumn,prx,showpacs, amsmath, amssymb, superscriptaddress,aps,groupedaddress]{revtex4-1}
\usepackage{graphicx}
\usepackage{epstopdf}
\usepackage{float}
\usepackage{amsmath}
\begin{document}
\title{Dynamic and spectral properties of transmission eigenchannels in random media}
\author{Zhou Shi and Azriel Z. Genack}
\affiliation{$^1$Department of Physics, Queens College of the City University of New York, Flushing, NY 11367, USA
\\$^2$Graduate Center of the City University of New York, New York, NY 10016, USA}
\date{\today}

\begin{abstract}
The eigenvalues of the transmission matrix provide the basis for a full description of the statistics of steady-state transmission and conductance. At the same time, the ability to excite the sample with the waveform of specific transmission eigenchannels allows for control over transmission. However, the nature of pulsed transmission of transmission eigenchannels and their spectral correlation, which would permit control of propagation in the time domain, has not been discussed. Here we report the dramatic variation of the dynamic properties of transmission with incident waveform. Computer simulations show that lower-transmission eigenchannels respond more promptly to an incident pulse and are correlated over a wide frequency range. We explain these results together with the puzzlingly large dynamic range of transmission eigenvalues in terms of the way quasi-normal modes of the medium combine to form specific transmission eigenchannels. Key factors are the closeness of the illuminating waves to resonance with the modes comprising an eigenchannel, their spectral range, and the interference between the modes. We demonstrate in microwave experiments that the modal characteristics of eigenchannels provide the optimum way efficiently excite specific modes of the medium. \footnote{Electronic address: genack@qc.edu}
\end{abstract}
\maketitle
Subject Areas: Condensed Matter Physics, Optics 

\section{{\bf Introduction}}
A monochromatic wave impinging on a static disordered sample is scrambled to produce a stable speckled intensity pattern within and beyond the sample. This pattern is the result of the interference of waves following innumerable possible trajectories leading to any point and is not statistically correlated with the incident waveform. Nonetheless, if the transmission matrix (TM) {\it t} is known, it is possible to exert a measure of control over the output waveform by manipulating the incident field \cite{2008b,2010c,2012f,2012g,2012a,2013g,2014c,2014d}. The elements of the TM are the field transmission coefficients $t_{ba}$ between incident and outgoing channels of the empty waveguide, {\it a} and {\it b}, respectively. The number of such channels in samples of constant cross section and reflecting sides is essentially the ratio of the cross sectional area $A$ and the square of the field coherence length, $N\sim A/(\lambda/2)^2$, where $\lambda$ is the wavelength in the medium surrounding the sample. By adjusting the phase of the incident channels so that all the transmitted channels are in phase at a selected point, a focused spot can be produced at that point with peak intensity $N$ times larger than the average intensity before focusing \cite{2007d,2008g,2010b,2010c,2011g,2011h,2011i,2012c,2012d,2012d,2013i}. 

The {\it N} transmission eigenchannels are paired sets of orthogonal incident and outgoing field patterns with associated transmission eigenvalues $\tau_n$ ranked from $n=1$ to {\it N} in order of decreasing transmission. As a result of the repulsion between the eigenvalues of the transmission matrix, they span the range from unity, $\tau_1=1$, to $\tau_N < \exp(-2L/\ell)$ \cite{1982a,1984a,1988c,1990g,1990f,1991c,1997c}, where $L$ is the sample length and $\ell$ is the transport mean free path in which the direction of a wave is randomized. The exponentially large dynamic range of transmission eigenvalues allows for the control of net transmission through the sample \cite{2008b,1984a,2012f,2012g,2014c,2014d}. However, there are practical limitations on the dynamic range of transmission because the complete transmission matrix of a sample cannot be easily determined \cite{2010h,2013j}.

The wide range of transmission eigenvalues was discovered by Dorokhov \cite{1982a,1984a} who considered the scaling of the conductance of wires. In units of the quantum of conductance, $e^2/h$, where $e$ is the electron charge and $h$ is Planck's constant, the conductance is equal to the transmittance, $T$, which is the sum of all flux transmission coefficients between channels, and can be expressed as the sum of all transmission eigenvalues, $T=\sum_{a,b=1}^N |t_{ba}|^2 = \sum_{n=1}^N \tau_n$ \cite{1970a,1981b,1981c}. Dorokhov \cite{1982a} found that a multichannel medium scales as a collection of $N$ independent eigenchannels, each with its own auxiliary localization length, $\xi_n$. The variation of $\tau_n$ with $L$ is given by  $\tau_n=1/\cosh^2(L/\xi_n)$. The inverse auxiliary localization lengths increase linearly with $n$ as $1/\xi_n=(2n-1)/2\xi$ for $n<N/2$ and somewhat faster for $n>N/2$ \cite{1990g,1991c,1997c}. Here, $\xi$ is the localization length. For diffusive samples with $L<\xi$, transmission is higher than $1/e$ for approximately ${\textsl g}=\langle T\rangle$ eigenchannels \cite{1984a,1986a}, of which the auxiliary localization lengths are greater than $L$. Here, $\langle \dots\rangle$ indicates the average over an ensemble of random sample configurations. The transmission eigenchannels are the $N$ pairs of incident and outgoing waveforms $V_n$ and $U_n$ corresponding to the $n$th column of the complex unitary matrices $U$ and $V$ found in the singular value decomposition of the TM, $t=U\Lambda V^\dagger$. $\Lambda$ is a diagonal matrix with elements $\lambda_n=\sqrt{\tau_n}$ \cite{1984a,1988c,1997c}. 

Since transmission eigenchannels are found from an analysis of the transmission matrix at a particular frequency, the focus of the studies of the TM has been on the transmission and control of monochromatic waves \cite{2012a}. There has also been interest in focusing in space and time by bringing all components of the transmitted field into phase at a particular point in space and time \cite{2011a,2011b,2011b1,2013b}. Generally, studies of pulse transmission and focusing have dealt with incident waves with a fixed incoming spatial waveform over the pulse bandwidth. For a short pulse with broad bandwidth, the ensemble average of the time of flight distribution is the Fourier transform of the field correlation function \cite{1990a}. For diffusive sample, the time of flight distribution of transmitted radiation is given by solving the diffusion equation describing the incoherent walk of photons. This could no longer be the case when the incident wave corresponds to a transmission eigenchannels since the properties of the eigenchannels reflects the coherence of the wave. In addition, the time and flight distribution and the field correlation function would no longer be a Fourier transform pair. At present, neither the transmitted pulse profile nor the field or intensity correlation functions for specific transmission eigenchannels is known. Until now, interest in the dynamics of particular transmission eigenchannels has been limited to the dwell time of the wave through the medium at a particular frequency \cite{2015a}. However, the dynamics of transmission eigenchannels are of great potential interest since transmission eigenchannels are orthogonal sets of waveforms on the sample's incident and outgoing surfaces that may be determined experimentally. Transmission eigenchannels may thereby provide a basis for controlling the dynamic and spectral properties of transmission \cite{2015c}. 

Because the incident waveforms $V_n$ change with frequency, pulsed transmission has generally been analyzed in terms of the quasi-normal modes of the medium with volume speckle patterns that are independent of frequency. Such quasi-normal modes are resonances of the open medium \cite{1998e}, which we shall refer to as ``modes" for simplicity. These are solutions of the wave equation and are analogous to levels in electronic systems. A possible approach to the description of the dynamics of transmission eigenchannels is to find the way they are composed of modes \cite{arxiv0}. This is clear with regard to the first transmission eigenchannel in transmission on resonance with spectrally isolated modes of localized waves. The speckle patterns of the first transmission eigenchannel and the resonant mode at the output surface of the sample are nearly identical \cite{2012f,2014f}. However, for both diffusive and localized waves, the way low transmission eigenchannels may be formed from the modes of the medium has not been clarified. In addition to enabling the control of pulsed transmission, the understanding of the modal makeup of transmission eigenchannels can also be exploited to control the excitation of modes within the sample. This would be of importance in applications such as low-threshold random lasing, which require deposition of energy in long-lived modes deep within the bulk of the medium \cite{1994c,1999d,2001d,2011f}. Relating transmission eigenvalues and modes in random media is also of fundamental interest. The close connection between these excitations is manifest in the equality for diffusive waves of two localization parameters, the dimensionless conductance {\textsl g} and the Thouless number $\delta$ \cite{1974b,1977a}, which is the ratio of the average spectral width and spacing of modes.
 
In this paper, we show that, in addition to differences in the magnitude of steady-state transmission, transmission eigenchannels have distinctive temporal and spectral properties. Low-transmission eigenchannels respond promptly to an incident pulse and their eigenvalues are correlated over a wide frequency range. We show how both dynamic and steady-state characteristics of transmission eigenchannels emerge from the way eigenchannels are made up of modes. Lower transmission eigenchannels are composed of a larger number of modes that are further from resonance and are consequently more strongly suppressed by destructive interference. This is conveniently formulated by expressing the TM as a superposition of modal TMs. We demonstrate in microwave experiments that the representation of channels in terms of modes makes it possible to excite specific modes of the medium. This can be exploited for controlling energy deposition within the medium and selectively exciting long-lived modes. 

\section{{\bf Simulations}}
We consider the static and dynamic properties of transmission eigenchannels in disordered quasi-one dimensional waveguides, which are locally two dimensional, by analyzing simulations of the TM at each frequency over a range of frequencies. Simulations are carried out in the crossover to localization in an ensemble of samples with ${\textsl g}=0.26$, in which a number of transmission eigenchannels and modes typically contribute to transmission. Though there is appreciable spectral overlap of transmission peaks and of modes for this ensemble, it is still possible to analyze field transmission spectra into a superposition of the modes of the medium \cite{2011j}. It is thereby possible to obtain their central frequencies, linewidths, and speckle patterns in transmission. This makes it possible to probe the modal basis of the steady-state and dynamic properties of transmission eigenchannels. Because the key to understanding the dynamics of transmission eigenchannels lies in the behavior of low-transmission eigenchannels, which are buried in the noise in experiments in random ensembles with ${\textsl g}<1$, simulations rather than measurements are utilized for the most part in this study. In contrast, experimental studies of the control energy within the medium for high transmission eigenchannels are not limited by measurement noise and such control is demonstrated in section III.c in microwave measurements of transmission in random dielectric waveguides for which ${\textsl g}=0.37$.

The wave equation, $\triangledown^2E(x,y)+k_0^2\epsilon(x,y)E(x,y)=0$, is discretized on a square grid and solved using the recursive Green's function method \cite{1991g}. The random position-dependent dielectric constant, $\epsilon(x,y)=1+ \Delta\epsilon(x,y)$ varies on the grid with $\Delta\epsilon(x,y)$ drawn from the rectangular distribution [-0.9,0.9]. Thus, the index of refraction of the empty waveguide before and after the random sample of unity and the averaged index of refraction of the random portion of the waveguide are matched and internal reflection is minimal. The wavelength of the incident wave in the leads is set to be $2\pi$ in units of the grid spacing at 10 GHz. The width of the waveguide is chosen so that the number of propagating channels in the empty waveguide is $N=16$. The statistics of transmission eigenchannels was studied in spectra between 10 and 10.06 GHz in 100 sample configurations. A modal analysis was carried out in 24 of these samples. The correlation frequency, which we take to be the half-width of the correlation function of the field correlation function, is 2 MHz.
\section{{\bf Results and discussion}}
\subsection{{\bf Dynamic and spectral properties of transmission eigenvalues}}
Pulses with Gaussian amplitude in the frequency and time domains are launched at the input of the sample with speckle patterns $V_n$ for the $n$th eigenchannel at the central frequency of the incident pulse. The pulses are formed by having all frequency components in phase at the time origin. The pulsed transmission of flux for incident eigenchannels $n=1,2,5$ are shown in Figs. 1a-c. The incident pulses are at a carrier frequency 10.0332 GHz corresponding to the vertical red dashed line in Figs. 1d-f showing the spectra of the transmission eigenvalues $\tau_n$ for the same eigenchannel indices {\it n}. 
\begin{figure}[htc]
\centering
\includegraphics[width=3.375in]{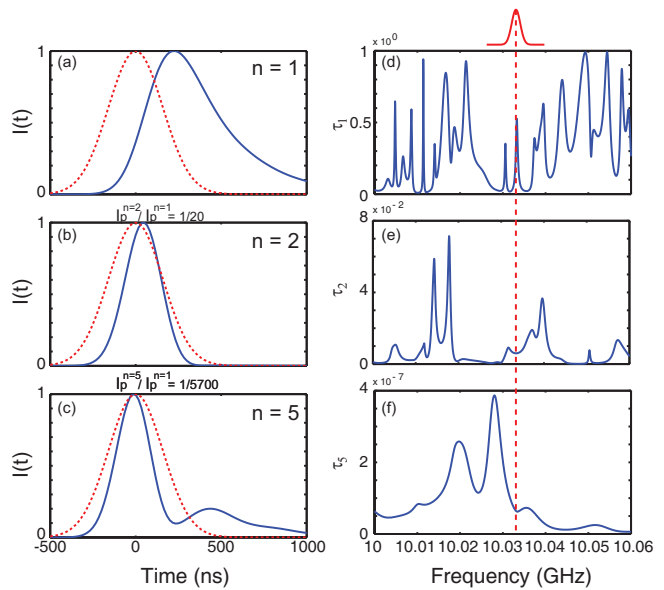}
\caption{(a-c) Simulation of responses (blue curves) to an incident Gaussian pulse (red dotted curves) with waveform $V_n$ of the transmission eigenchannels $n=1,2,5$ for the central frequency of the pulse in a single random configuration for a sample with $N=16$ channels drawn from an ensemble with {\textsl g}=0.26. The central frequency of the pulse is indicated by the red dashed line in the frames. The peak of the response moves towards the peak of the incident pulse as the net transmitted flux decreases and $n$ increases. (d-f) Spectra of transmission eigenchannels $\tau_n$ for $n=1,2,5$. The intensity of the incident Gaussian pulse in the frequency domain is shown as the red curve centered above the red dashed line. The width of the spectral features increases as the transmission decreases and $n$ increases.} \label{Fig1}
\end{figure} 
The peak in the transmitted pulse in Figs. 1a-c is seen to shift towards earlier times as $n$ increases and the total energy and average dwell time of the transmitted pulses decreases. The maximum intensity in the transmitted pulse decreases significantly for large value of $n$ as indicated by the decreasing ratio of the peak intensity relative to the peak for the first transmission eigenchannel in Figs. 1b and 1c. The ensemble averaged value of the delay of the first peak in the transmitted pulse for different incident eigenchannels is presented in Figs. 2a. In addition to the changes found in pulsed transmission, a systematic change is seen in the character of spectra of $\tau_n$ in Figs. 1d-f with the spectral features broadening with increasing $n$. This is manifest in the broadening of the cumulant correlation function with frequency shift of the transmission eigenvalues normalized by their average values, $C_{\tau_n}=\langle \tau(\Delta \nu)\tau\rangle/\langle \tau\rangle^2-1$ seen in Fig. 2b. This is in contrast to the single intensity correlation functions \cite{1987a} for fixed incident waveforms, which reflect the average dynamics of the medium. A plot of the correlation frequency for various eigenchannels is given in the inset of Fig. 2b. 
\begin{figure}[htc]
\centering
\includegraphics[width=3.375in]{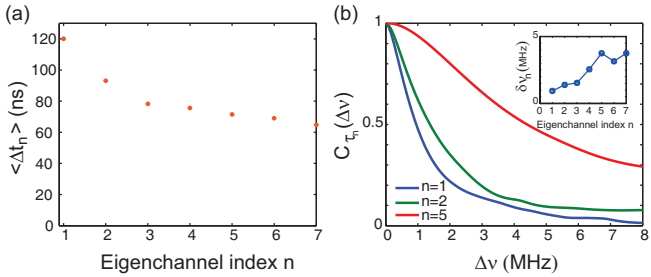}
\caption{(a) Variation of average delay of peak of transmitted pulse with incident waveform of the $n$th transmission eigenchannel at the center of the pulse vs. eigenchannel index $n$ in an ensemble with {\textsl g}=0.26. (b) Normalized cumulant correlation function of transmission eigenchannel $\tau_n$ vs. frequency shift. The inset shows the increase of the correlation frequency with $n$. The correlation frequency is taken as the frequency shift at which the correlation function falls to half its value.} \label{Fig2}
\end{figure} 
We found that, along with the changes in the spectral correlation function of the magnitude of transmission of the eigenchannels, the correlation of their field speckle patterns changes. The correlation function of the square of the corresponding normalized speckle patterns $U_n$, $C_{U_n}(\Delta\nu)=|\langle (U_n^*(\nu),U_n(\nu+\Delta\nu))\rangle|^2$ in which $({\bf x}, {\bf y})$ indicates the inner product of two vectors, are seen in Fig. 3 to broaden as $n$ increases. 
\begin{figure}[htc]
\centering
\includegraphics[width=3.375in]{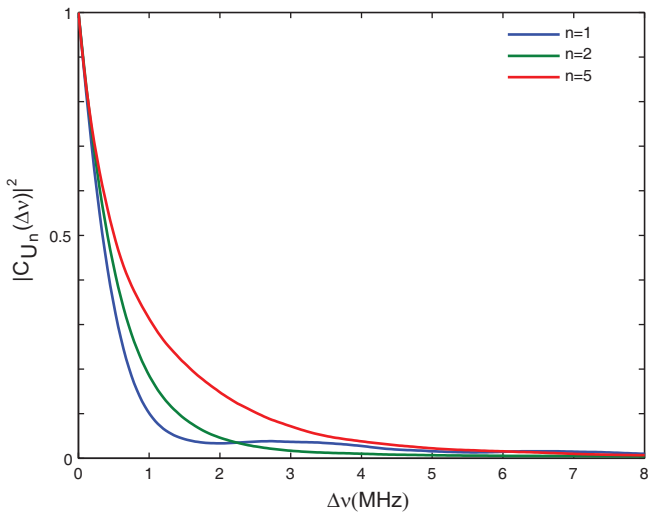}
\caption{Normalized correlation function of the square of the overlap integral of the waveform of the transmission eigenchannel on the output surface, $U_n$, with the waveform of $U_n$ at a shifted frequency $\Delta\nu$.} \label{Fig3}
\end{figure} 

The falling transmission and growing correlation frequency of eigenchannel correlation functions for lower transmission eigenchannels suggests that lower-transmission eigenchannels may be composed of modes that are spectrally more remote from the excitation frequency. For a given mode, with Lorentzian line normalized to unity at the line center, $|\phi^m(\nu-\nu_m)|^2=|\frac{\Gamma_m/2}{\Gamma_m/2+1i(\nu-\nu_m)}|^2$, where $\phi^m(\nu-\nu_m)$ is the response of the field associated with the mode. Here, $\nu$ is the frequency of the incident wave, and $\nu_m$ and $\Gamma_m$ are the central frequency and linewidth of the mode, the fractional change in intensity falls more slowly with frequency shift as the shift, $\nu-\nu_m$, increases. Similarly, the phase derivative with frequency shift associated with the mode, which is related to the delay time is smaller and falls more slowly with increasing frequency shift from $\nu_m$. 
\begin{figure}[htc]
\centering
\includegraphics[width=3.375in]{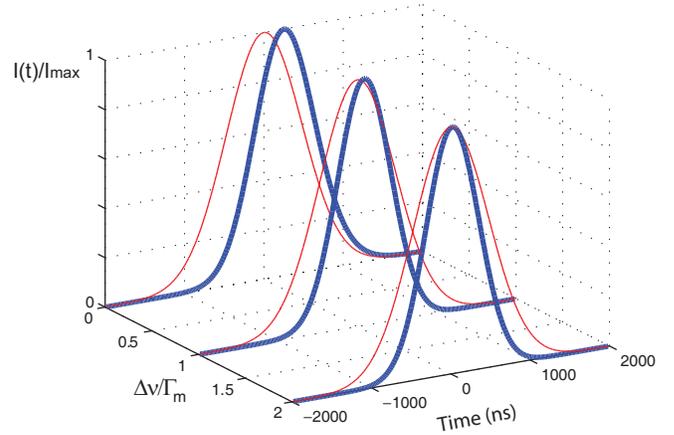}
\caption{Response of a Lorentzian mode to excitation by a Gaussian pulse for resonant excitation and for the central frequency of the pulse shifted one and two linewidths from the center of a Lorentzian line with linewidth $\Gamma_m$. Bandwidth $\sigma$ of the Gaussian pulse is 1/3 of the linewidth $\Gamma_m$ of the Lorentzian mode. The peak in the response is seen  to move towards the time of the incident pulse.} \label{Fig4}
\end{figure} 
The dynamic response of lower transmission eigenchannels bears a similarity to systematic changes in pulsed transmission as the excitation frequency is shifted further off resonance. In Fig. 4, we consider the changing dynamic response to an exciting pulse which is tuned away from resonance with a single mode. The modal response for excitation on-resonance is delayed relative to the incident pulse, while the response becomes more prompt as the excitation is shifted off resonance, with the peak of the transmitted pulse nearly coinciding with that of the incident pulse. This trend is similar to the changing character of transmission through a random medium with increasing eigenchannel index $n$ shown in Figs. 1a-c. The increasingly prompt response to an incident pulse of radiation as the pulse is shifted off resonance is also exhibited in two level systems such as nuclear spins or electronic transitions in atoms in nuclear magnetic or optical nutation \cite{1949a,1968b,1971a}. The pulsed response in these cases is explained in terms of the nutation of the magnetization or of a generalized optical dipole moment about the effective field in the frame of reference rotating at the excitation frequency. Far off-resonance, the effective driving field and consequentially the Rabi frequency is large so that the magnetization or effective polarization rapidly follows the exciting pulse.

\subsection{Modal basis for transmission eigenchannels}
To explore the dynamical and spectral characteristics of transmission eigenchannels and the wide range of transmission eigenvalues, we express the TM as a superposition of contributions of the modes as,  $t(\nu)=\sum_{m} t^m\phi^m(\nu)$, where $t^m$ is the TM of the $m$th mode on resonance and   gives frequency variation of the field associated with the mode. The elements $t^m_{ba}$ of $t^m$ describe the coupling from incident channel {\it a} into the $m$th mode of the open system and then coupling into output channel {\it b} \cite{1991i,2009c}. The modal TM can be expressed via the singular value decomposition as $t^m=u^m\lambda^mv^{\dagger m}$. The transmitted flux for excitation by the $n$th column of $v^m$ is $(\lambda_n^m)^2=\tau_n^m$. We find small changes in the output speckle patterns of modes in the analysis of transmitted field spectra into modes. This may be an inherent property of open systems or due to an inadequacy of the modal analysis. As a result, the columns in the modal TM are not strictly proportional, though they are highly correlated and $t^m$ is nearly of rank one. $\lambda_{2}^m$ of the $t^m$ determined from fitting field spectra to the sum of modes, does not vanish, but is typically small with $\langle\lambda_{2}^m/\lambda_{1}^m \rangle\sim 0.1$, while $\langle\tau_2^m/\tau_1^m\rangle\sim 0.04$. Nonetheless, for the present purpose of exploring the qualitative properties of transmission eigenchannels, the modal TM can be approximated as, $t^m=u_1^m\lambda_1^mv_1^{\dagger m}$, where $v_1^m$ and $u_1^m$ are the incoming and outgoing field patterns of the first transmission eigenchannel for the mode. Therefore a particular mode is most effectively excited by illuminating the sample with the incident pattern, $v_1^m$ with transmission $\tau_1^m=(\lambda_1^m)^2$.

Associating low transmission eigenchannels with off-resonance modes is in keeping with the understanding that discrete peaks in transmission in deeply localized media correspond to being on resonance with a mode of the medium. More generally, however, the number of modes which individually contribute to transmission at a level greater than $\tau_N$ is much greater than the number of channels $N$. The degree of excitation of modes falls with frequency shift as a Lorentzian, with an inverse square decay which is slow compared to the exponential falloff of transmission eigenvalues with increasing $n$. Thus the remoteness of a transmission eigenchannel from resonance with modes does not by itself explain the small values of transmission eigenvalues for high indices $n$.

To investigate the modal makeup of the transmission eigenchannels, we consider the correlation function between the normalized output field patterns of the $m$th mode, $u_1^m$, and the $n$th eigenchannel, $U_n$, for a wave with frequency $\nu$ shifted from the central frequency of the mode by $\Delta\nu=\nu-\nu_m$, $C_n^m(\Delta\nu)=(U_n^*,u_1^m)$. We see in Fig. 5 that the speckle pattern in transmission of the first transmission eigenchannel is strongly correlated with the closest modes and more weakly correlated with modes with central frequencies shifted further from resonance. At the same time, resonant modes are increasingly anti-correlated with eigenchannels with increasing index n at the excitation frequency. The peak in the correlation function is seen to shift and broaden with increasing $n$. Because simulations are carried out in a sample with $N=16$, the degree of correlation at large frequency shifts at which the speckle patterns of modes are statistically independent, does not vanish and approaches the value $N^{-1}$. 
\begin{figure}[htc]
\centering
\includegraphics[width=3.375in]{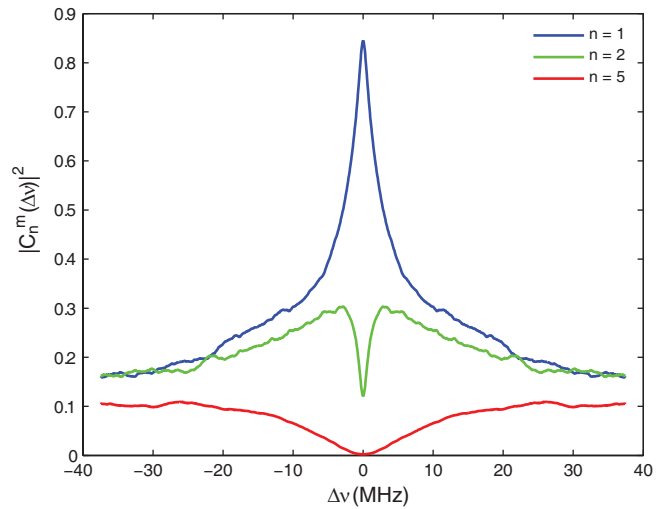}
\caption{Spectral correlation of field patterns modes and eigenchannels. The square of the correlation function $C_n^m(\Delta\nu)=(U_n^*,u_1^m)$ between the normalized output field pattern of the $m$th mode, $u_1^m$, and the $n$th eigenchannel, $U_n$, for a wave shifted from the central frequency of the mode by $\Delta\nu=\nu=\nu_m$.} \label{Fig5}
\end{figure} 

The way that modes are superposed to give $\tau_n$ can be explained in terms of the decomposition of the TM into the sum of modal TMs. The various factors involved can be appreciated by considering them sequentially following Fig. 6. 
\begin{figure}[htc]
\centering
\includegraphics[width=3.375in]{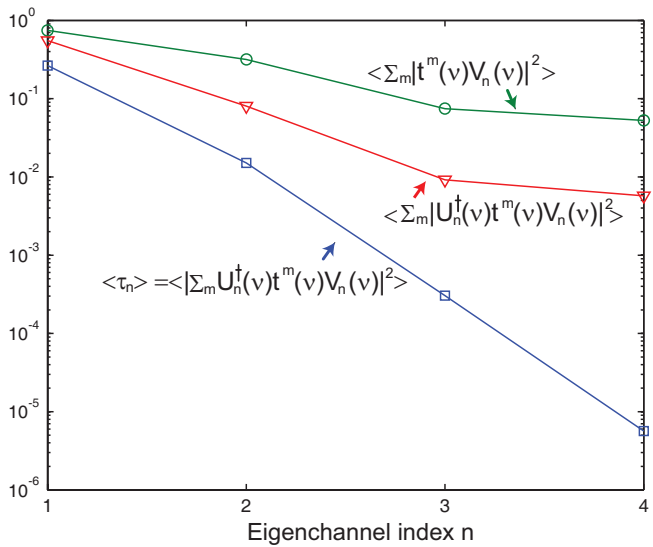}
\caption{The factors that taken together yield the transmission eigenvalues for the first four eigenchannels are shown. The upper set of green circles gives the average of the incoherent sum of transmitted flux in all modes excited by an incident waveform of a transmission eigenchannel, $V_n$. The middle set of red triangles is the average of the incoherent sum of modal contributions to transmission for incident wave $V_n$ with the output wave projected onto the 2D space of $U_n$ and $iU_n$.  The lower set of blue square gives the flux associated with the projections onto $U_n$ from all modes and is equal to $\tau_n$.} \label{Fig6}
\end{figure} 
The upper set of green circles in Fig. 6a gives the average of the incoherent sum of flux in all modes in the frequency range studied when the sample is excited by the eigenchannels $V_n$ for $n=1-4$. This may be written as, $\sum_{m} |t^m(\nu)V_n(\nu)|^2$. Next, we consider the average of the incoherent sum of modal contributions to transmission for the incident wave $V_n$ with the component of the transmitted wave projected onto the space of output eigenchannel $U_n$, $\sum_{m} |U_n^\dagger t^m(\nu)V_n(\nu)|^2$. This is shown as the middle set of red triangles in Fig. 6a. Transmission is further reduced by the random phasing of the projections onto $U_n$ from all modes. This can be visualized using a vector model in the two-dimensional space of speckle patterns which are the in- and out-of-phase components of $U_n$, $U_n$ and $iU_n$. This is shown in Fig. 7 for eigenchannel $n=1$ and 3 at a specific frequency. 
\begin{figure}[htc]
\centering
\includegraphics[width=3.375in]{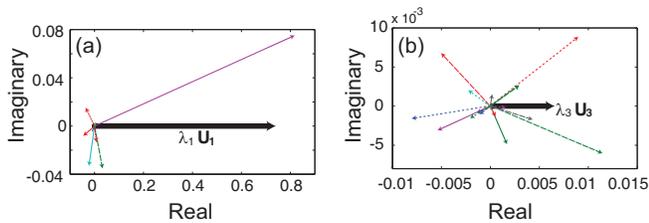}
\caption{The projection from the 2D space with unit vectors in- and out-of-phase with $U_1$ onto the $U_1$-axis for the incident wave of the first transmission eigenchannel $V_1$ in a single sample configuration.at a particular frequency. The resultant is $\lambda_1U_1$. The same as (a) but for the third transmission eigenchannel with incident wave $V_3$ and projection is onto the $U_3$-axis. Many modes with vectors of random phase contribute.} \label{Fig6}
\end{figure} 
The square of the length of the resultant vector, which lies along $U_n$, is equal to the transmission eigenvalue $\tau_n$. The resultant for $n=1$ is seen in Fig. 7a to be composed largely of a single vector. In contrast, for $n = 3$, the resultant is formed by the superposition of many vectors with comparable magnitudes and with angles distributed over $2\pi$ radiance. The resultant is even shorter than the length of several of the individual contributions. This illustrates the strong suppression of lower transmission eigenchannels by destructive interference of waves associated with different modes of the medium,  and provides a physical model for the small transmission eigenchannels found by Dorokhov. 

\subsection{Microwave measurements of the selective excitation of modes}
In addition to affording a qualitative description of pulse transmission and the large dynamical range of the transmission eigenvalues, the modal TM may be exploited to efficiently excite a specific mode in pulsed and steady-state transmission. We demonstrate this in microwave measurements of the spectrum of the TM carried out in collections of randomly positioned alumina spheres, in which the high transmission eigenchannels can be determined with good accuracy \cite{2012f}. The spheres of diameter 0.95 cm and refractive index 3.14 are embedded in Styrofoam shells to give an alumina sphere volume fraction of 0.068. Measurements are carried out in the frequency range just above the first Mie resonance of the spheres at 10 GHz in which scattering is strong. The spheres are contained within a section of a copper tube region of length $L= 40$ cm and diameter 7.3 cm. The field transmission coefficients are measured between points on the incident and output cross sections of the tube with use of a vector network analyzer. The source and detection antennas are the exposed 4-mm-long central conductors of a microwave cable. The antennas are moved over a grid of 49 points using translation stages. In the frequency range 10.13-10.18 GHz, the empty portion of the waveguide supports $N = 30$ transverse propagation modes. The sample is drawn from an ensemble of samples for which {\textsl g}=0.37. The impact of absorption on the statistics of transmission is removed by Fourier transforming the field spectra into the time domain then multiplying the time signal by $\exp(t/2\tau_a)$, and finally, Fourier transforming back into the frequency domain \cite{2012f}. Here $t$ is the delay from the peak of the incident pulse and $1/\tau_a$ is the absorption rate. The transmission eigenvalues for the TM corrected for absorption are summed to give the transmittance spectrum shown in Fig. 8a. 

\begin{figure}[htc]
\centering
\includegraphics[width=3.375in]{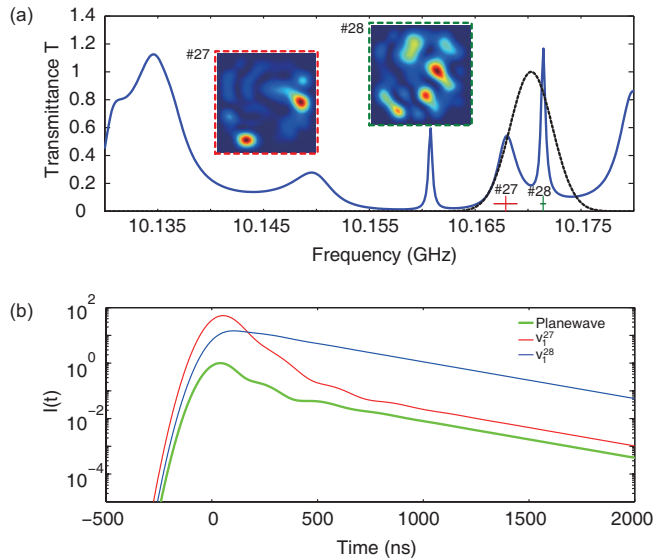}
\caption{The spectrum of transmittance in a single configuration of an ensemble with {\textsl g} = 0.37 from microwave experiment. The intensity spectrum of the incident Gaussian pulse is also shown. The central frequencies of the two modes that fall within the linewidth of the Gaussian pulse are shown as vertical lines and the horizontal lines indicate the corresponding linewidths. The intensity speckle patterns of the two modes for mode 27 and 28 at the output surface are shown in the green and red boxes. (b) Time dependence of the transmitted pulse for the incident Gaussian pulse with intensity spectrum shown in (a) for incident waveforms that area the, in order of transmitted energy are the first transmission eigenchannels for modes 28 and 27 and for a plane wave. The single exponential decay of the high-transmission pulse with incident waveform $v_1^{28}$ indicates that mode 28 is selectively excited.} \label{Fig7}
\end{figure} 
The degree to which a single mode can be selectively excited by an incident pulse with a bandwidth that overlaps several modes with different decay rates can be seen in the decay of the transmitted pulse. To evaluate the selective excitation of modes, we first find the central frequency and linewidth of the modes. This is accomplished by the simultaneous fit of the modal expansion of field transmission spectra with a common set of $\nu_m$ and $\Gamma_m$ for all pairs of source and detector positions. The central frequencies and linewidths of the modes are indicated in Fig. 8a. Once the set of $\nu_m$ and $\Gamma_m$ are found, the modal TMs $t^m$ can be found from the decomposition of the spectrum of the TM into a sum over modal TMs.

The variation in time of the transmitted flux due to a Gaussian pulse multiplying various incident waveforms is shown in Fig. 8b. The carrier frequency of the incident pulse of 10.1703 GHz lies between the central frequencies of the 27th and 28th modes of 10.1679 and 10.1714 GHz. The transmitted pulse is determined by the time dependent transmitted fields, which are obtained from the Fourier transform of the product of the Gaussian in the frequency domain as the dashed black curve, shown in Fig. 8a, and the corresponding field transmission spectrum. The incident patterns of the pulses in Fig. 8b are a plane wave and the waveforms corresponding to the first eigenchannels of the TM for modes 27 and 28 at the central frequencies of these modes, $\nu_{27}$ and $\nu_{28}$. The decay rates of these modes are $2\pi\Gamma_{27}=1.4\times10^7 s^{-1}$ and $2\pi\Gamma_{28}= 3.1\times10^6 s^{-1}$. At long time, transmission for the various excitation waveforms decays at the rate of 28th modes which is the longest-lived of the modes excited by the pulse. For the incident waveform $v_1^{28}$, the wave is seen to decay from its peak value with a constant decay rate $2\pi\Gamma_{28}$, indicating that the 28th mode is selectively excited. For incident waveform $v_1^{27}$, most of the energy decays at the higher decay rate of the 27th mode at earlier times. In contrast, for plane-wave illumination, the oscillation of transmission at early times indicates the beating between two excited modes. In addition, the transmission averaged over time for the plane wave is well below the level for the incident waves that specifically excite one or another of the two spectrally close modes. This is because though all transmission eigenchannels are approximately equally represented in an incident plane wave, the lower transmission eigenchannels do not contribute substantially to transmission. The ratio of the integrated flux for a pulse with carrier frequency at $\nu_{28}$ with incident waveforms of $v_1^{28}$ and a plane wave is 34. 

Individual modes can be selected when either the correlation between the field speckle patterns of modes is small or when the modes are separated by at least a linewidth of the broader mode. When these conditions do not hold, however, the first transmission eigenchannels for the modes may be similar and it may no longer be possible to excite a single mode via illumination on the incident surface. Indeed, the selective excitation of modes in these circumstances can violate the principle of the conservation of energy. We show in Fig. 9, that the first singular value of $t^m$, and hence $\tau^m$, may exceed unity, when illuminating the sample with the first transmission eigenchannel at the central frequency of mode \#7 10.0215 GHz as indicated by the dashed line. Since the two modes primarily excited overlap spectrally and the field speckle patterns associated interfere destructively, the overall transmission is below unity. This is seen in the vector diagram in Fig. 9b, where the contribution to $\lambda_1U_1$ from these two modes has the opposite sign and so leads to a small resultant vector. 
\begin{figure}[htc]
\centering
\includegraphics[width=3.375in]{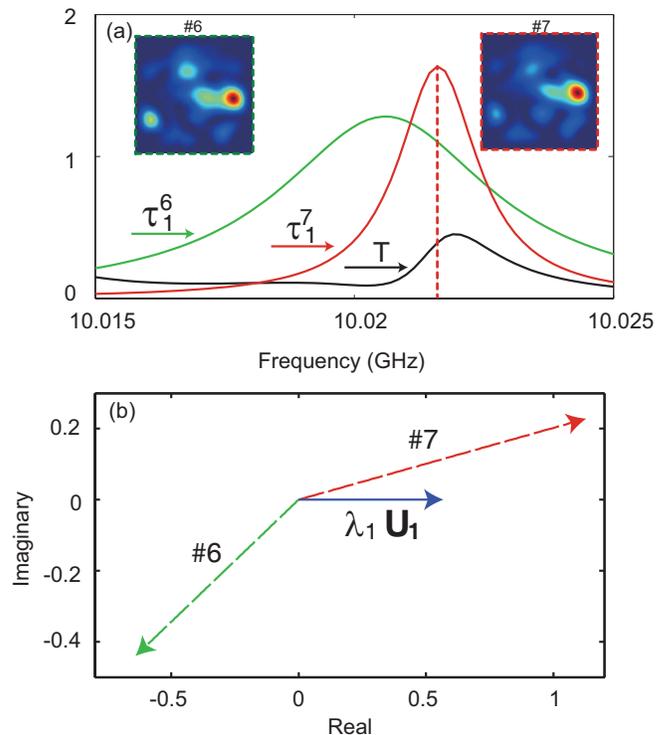}
\caption{(a) Spectra of the transmittance and the first transmission eigenvalues for modes 7 and 8 in one random sample configuration. The overlapping modes are seen to have similar speckle patterns. (b) The vector diagram in the space ($U_1 , iU_1$) for the first transmission eigenchannel at the central frequency of the Mode 7 is shown. The first modal singular value for mode 7 alone is seen to correspond to a transmission coefficient greater than unity. Because the speckle patterns of modes 7 and 8 are similar, the modes cannot be selectively excited and transmission is therefore smaller than unity.} \label{Fig8}
\end{figure} 

\section{Conclusion}
In summary, we have shown that along with the well-known strong variation in transmission eigenvalues with eigenchannel index {\it n}, dramatic changes emerge in their temporal and spectral characteristics. These changes point the way to the origin of the puzzling large range of transmission eigenvalues and provide the basis for the efficiently exciting a medium in specific modes.

An incident Gaussian pulse with the waveform of the $n$th transmission eigenchannel at the central frequency of the pulse rises earlier with a peak in transmission that shifts towards the center of the pulse as $n$ increases. This makes it possible to extend the realm over which transmission can be controlled from the steady-state to the time domain. In line with the changes in the transmitted pulse with increasing $n$, the spectral correlation frequency of lower transmission eigenchannels broadens significantly. We also find that the speckle pattern of the highest transmission eigenchannel is correlated while the patterns of low transmission eigenchannels are anti-correlated with modes resonant with the incident wave. These results show that the lower the transmission eigenchannel, the greater is the frequency shift of the modes forming them from the excitation frequency. These changes in transmission in the time and frequency domains arise because lower-transmission eigenhannel are made up of modes that are further from resonance with the incident wave. The explanation of the exponentially small values of steady-state and pulsed transmission requires an additional factor-the increasingly potent destructive interference among the larger number of modes contributing to lower transmission eigenchannels.

These findings give a qualitative understanding of the modal make up of transmission eigenchannels and of the dynamic and spectral properties of transmission eigenchannels. A theoretical treatment of the spectral correlation of eigenchannel waveforms \cite{arxiv0,arxiv1} and their correlation with modal waveforms are of importance in controlling broadband transmission. The variation of the the minimum reflection eigenchannel with gain has recently been discussed in terms of their modal makeup \cite{arxiv2}. Expressions for these quantities would make it possible to more reliably anticipate the degree to which modes can be selected within a medium and transmitted pulses can be shaped. 

The descriptions of pulsed transmission in terms of transmission eigenchannels and modes differ in many respects. The incident waveform of a transmission eigenchannel $V_n$ changes with frequency, while modal waveforms are independent of frequency. However, in practice, the waveform of incident pulses is fixed over the bandwidth of the pulse so that more than a single transmission eigenchannel is excited by an incident pulse. Low-transmission eigenchannels are correlated over a wide frequency range and a short broadband pulse may excite a single eigenchannel but a large number of modes. On the other hand, for high-transmission eigenchannels in samples with spectrally overlapping modes, a pulse with fixed incident waveform and with spectral width greater than the correlation frequency of the eigenchannel may excite several eigenchannels a small number of modes over the frequency range of the pulse. Another difference between the eigenchannel and mode descriptions of transmission is that eigenchannels are a complete orthogonal set over the sample surface, while the field patterns of quasi-normal modes over a cross section are not orthogonal.  We find in simulations and measurements that the output speckle patterns of modes vary slightly with incident channel. As a result, modal TMs are not of rank one. The impact of the openness \cite{2005d} of the sample on the modal analysis of the transmission eigenchannels is therefore of great interest and requires further study. 

The description of eigenchannels in terms of modes points the way towards efficient pulsed excitation of a medium and provides a basis for selecting specific modes. This selectivity is optimized in pulsed excitation for resonant illumination with a spectrum that overlaps that of the mode and a waveform that matches the modal field speckle pattern. Such excitation of long-lived modes could enhance nonlinear effects and lower the threshold for random lasers.

\section{Acknowledgment}
We thank Matthieu Davy and Li Ge for discussions and Arthur Goetschy and A. Douglas Stone for providing the simulation code to calculate the transmission matrix through a two dimensional disordered waveguide. This work was supported by the National Science Foundation through Grant No. DMR-1207446.

\end{document}